\documentclass[aps,prl,reprint,superscriptaddress,amsmath,amssymb]{revtex4-1}

\usepackage{microtype}
\usepackage{xcolor}
\usepackage{graphicx}
\usepackage{dcolumn}
\usepackage{multirow}
\usepackage{siunitx}
\usepackage{soul}
\DeclareSIUnit\Dq{Dq}
\DeclareSIUnit\bohrmagneton{\ensuremath{\mu_{\mathrm{B}}}}
\DeclareSIUnit\entropy{\ensuremath{J/mol-Ru\cdot K}}
\DeclareSIUnit\molaremu{\ensuremath{emu/mol-Ru}}
\DeclareSIUnit\wtpercent{\ensuremath{wt.\%}}
\usepackage[colorlinks]{hyperref}
\usepackage{braket}
\usepackage{amsmath}
\newcommand*\up{\mathord{\uparrow}}
\newcommand*\dn{\mathord{\downarrow}}

\begin{document}
	
\title{Formation of orbital molecules on a pyrochlore lattice induced by $A$-O bond covalency}

\author{A. Krajewska}
\affiliation{Max Planck Institute for Solid State Research, 70569 Stuttgart, Germany}
\affiliation{Institute for Functional Matter and Quantum Technologies, University of Stuttgart, 70550 Stuttgart, Germany}
\affiliation{ISIS Neutron and Muon Source, STFC Rutherford Appleton Laboratory, Chilton, Didcot, Oxon OX11 0QX, United Kingdom}
\author{A. Yaresko}
\affiliation{Max Planck Institute for Solid State Research, 70569 Stuttgart, Germany}
\author{J. Nuss}
\affiliation{Max Planck Institute for Solid State Research, 70569 Stuttgart, Germany}
\author{A. S. Gibbs}
\affiliation{ISIS Neutron and Muon Source, STFC Rutherford Appleton Laboratory, Chilton, Didcot, Oxon OX11 0QX, United Kingdom}
\author{S. Bette}
\affiliation{Max Planck Institute for Solid State Research, 70569 Stuttgart, Germany}
\author{M. Blankenhorn}
\affiliation{Max Planck Institute for Solid State Research, 70569 Stuttgart, Germany}
\affiliation{Institute for Functional Matter and Quantum Technologies, University of Stuttgart, 70550 Stuttgart, Germany}
\author{R. E. Dinnebier}
\affiliation{Max Planck Institute for Solid State Research, 70569 Stuttgart, Germany}
\author{D. P. Sari}
\affiliation{Graduate School of Engineering and Science, Shibaura Institute of Technology}
\affiliation{Meson Science Laboratory, Nishina Center for Accelerator-Based Science, RIKEN, 2-1 Hirosawa, Wako 351-0198, Japan}
\author{I. Watanabe}
\affiliation{Meson Science Laboratory, Nishina Center for Accelerator-Based Science, RIKEN, 2-1 Hirosawa, Wako 351-0198, Japan}
\author{J. Bertinshaw}
\affiliation{Max Planck Institute for Solid State Research, 70569 Stuttgart, Germany}
\author{H. Gretarsson}
\affiliation{Max Planck Institute for Solid State Research, 70569 Stuttgart, Germany}
\affiliation{Deutsches Elektronen-Synchrotron DESY, Notkestr. 85, D-22607 Hamburg, Germany}
\author{K. Ishii}
\affiliation{Synchrotron Radiation Research Center, National Institutes for Quantum Science and Technology, Hyogo 679-5148, Japan}
\author{D. Matsumura}
\affiliation{Materials Sciences Research Center, Japan Atomic Energy Agency, Hyogo 679-5148, Japan}
\author{T. Tsuji}
\affiliation{Materials Sciences Research Center, Japan Atomic Energy Agency, Hyogo 679-5148, Japan}
\author{M. Isobe}
\affiliation{Max Planck Institute for Solid State Research, 70569 Stuttgart, Germany}
\author{B. Keimer}
\affiliation{Max Planck Institute for Solid State Research, 70569 Stuttgart, Germany}
\author{H. Takagi}
\affiliation{Max Planck Institute for Solid State Research, 70569 Stuttgart, Germany}
\affiliation{Institute for Functional Matter and Quantum Technologies, University of Stuttgart, 70550 Stuttgart, Germany}
\affiliation{Department of Physics, University of Tokyo, Tokyo 113-0033, Japan}
\author{T. Takayama}
\affiliation{Max Planck Institute for Solid State Research, 70569 Stuttgart, Germany}
\affiliation{Institute for Functional Matter and Quantum Technologies, University of Stuttgart, 70550 Stuttgart, Germany}

\begin{abstract}
The pyrochlore ruthenate In$_2$Ru$_2$O$_7$ displays a subtle competition between spin-orbital entanglement and molecular orbital formation. At room temperature, a spin-orbit-entangled singlet state was identified. With decreasing temperature, In$_2$Ru$_2$O$_7$ undergoes multiple structural transitions and eventually forms a nonmagnetic ground state with semi-isolated Ru$_2$O units on the pyrochlore lattice. The dominant hopping through the Ru-O-Ru linkage leads to molecular orbital formation within the Ru$_2$O units. This molecular orbital formation is unique in that it involves the O$^{2-}$ anions, unlike the transition-metal dimers observed in systems with edge-sharing octahedra. We argue that the covalent character of In-O bonds plays a pivotal role in the structural transitions and molecular orbital formation and such bonding character of ``$A$-site'' ions is an important ingredient for electronic phase competition in complex transition metal oxides.
\end{abstract}


\maketitle

Complex transition-metal oxides are a platform for a plethora of exotic electronic phases and functions where multiple degrees of freedom of $d$-electrons, together with underlying lattice topology, are at play. The ground states of these systems are governed by a subtle balance of the relevant electronic parameters such as Coulomb repulsion, bandwidth, and crystal field. 4$d$ ruthenium compounds have so far played a remarkable role in providing such exotic phases, encompassing unconventional superconductivity~\cite{Maeno1994}, metal-insulator transition~\cite{Nakatsuji2000,Yamamoto2007} and quantum magnetism~\cite{Plumb2014}.

Besides metallic and magnetic insulator ground states, some ruthenium compounds exhibit a nonmagnetic and insulating state accompanied by the formation of molecular orbitals comprising $d$-electrons. A prominent example is honeycomb ruthenate Li$_2$RuO$_3$, which undergoes a dimerization of the Ru atoms below $\sim \SI{550}{\kelvin}$~\cite{Miura2007,Miura2009}, where 4$d$ electrons are accommodated into the molecular orbitals localized on the dimers. A similar dimerization has been found in 4$d$ and 5$d$-based honeycomb systems, especially under high pressures~\cite{Bastien2018,Hermann2019}. The spatially extended $t_{2g}$ orbitals in the dimers have substantial overlap across the edges of the anion octahedra and split into bonding and antibonding molecular orbitals. Much more complex orbital molecules may be formed on a frustrated lattice, where a simple arrangement of dimers may be unfavourable~\cite{Pen1997,Radaelli2005, Browne2017}.

In heavy transition-metal compounds such as ruthenates, another key ingredient for their electronic properties is spin-orbit coupling. In those materials, the magnitude of spin-orbit coupling is non-negligible and comparable to the other electronic parameters such as Hund’s coupling and non-cubic crystal field, often yielding spin-orbit-entangled $J_{\mathrm{eff}}$ states~\cite{Takayama2021}. In ruthenium compounds, Ru$^{4+}$ ions ($t_{2g}^4$ configuration), most commonly seen in complex ruthenium oxides, may give rise to a spin-orbit-entangled $J_{\mathrm{eff}} = 0$ singlet state. While the $J_{\mathrm{eff}} = 0$ state is non-magnetic, exotic magnetic ground states are expected to emerge owing to exchange interactions through the upper-lying $J_{\mathrm{eff}} = 1$ triplet separated by the spin-orbit coupling. It has been proposed that if the gap between the $J_{\mathrm{eff}} = 0$ and 1 states is small compared to the exchange interaction, magnetic ordering can occur, which is viewed as the condensation of the $J_{\mathrm{eff}} = 1$ triplet and dubbed excitonic magnetism~\cite{Khaliullin2013}. Indeed, a layered perovskite Ca$_2$RuO$_4$ has been demonstrated to display excitonic magnetism, where an amplitude mode with soft-moments, a hallmark of excitonic magnetism, has been identified~\cite{Jain2017}. The presence of excitonic magnetism remains unexplored in other ruthenates.

The competition of electronic phases including molecular orbital formation and spin-orbital entanglement is expected to be more pronounced in ruthenates with a frustrated lattice where conventional spin and orbital ordering may be suppressed. We have therefore focused on pyrochlore ruthenates $A_2$Ru$_2$O$_7$ ($A$: trivalent ion). The pyrochlore ruthenates have been regarded as $S = 1$ Mott insulators, possibly due to the presence of a strong trigonal distortion which may lift the degeneracy of the $t_{2g}$ orbitals and thus competes with spin-orbit coupling. While most of these pyrochlores show long-range magnetic ordering at low temperatures~\cite{Gardner2010}, Tl$_2$Ru$_2$O$_7$ exhibits a metal to nonmagnetic insulator transition at $\sim \SI{120}{\kelvin}$ ~\cite{Takeda1998,Lee2001}. The nonmagnetic singlet state has been attributed to the formation of a Haldane gap in the one-dimensional zigzag chains of $S$ = 1 Ru ions on the pyrochlore lattice~\cite{Lee2006}. The distinct behavior of Tl$_2$Ru$_2$O$_7$ may be related with the covalency of Tl-O bonds which has been discussed to play a role in the metal-insulator transition~\cite{Takeda1998,Lee2001,Ishii2000}. The covalent character of $A$-O bonds thus may be an important factor for the ground state of pyrochlore oxides. Additionally, the role of spin-orbit coupling has not been fully investigated in pyrochlore ruthenates.

\begin{figure*}[t]
	\centering
	\includegraphics[width=1\linewidth,trim=0 190 410 0,clip]{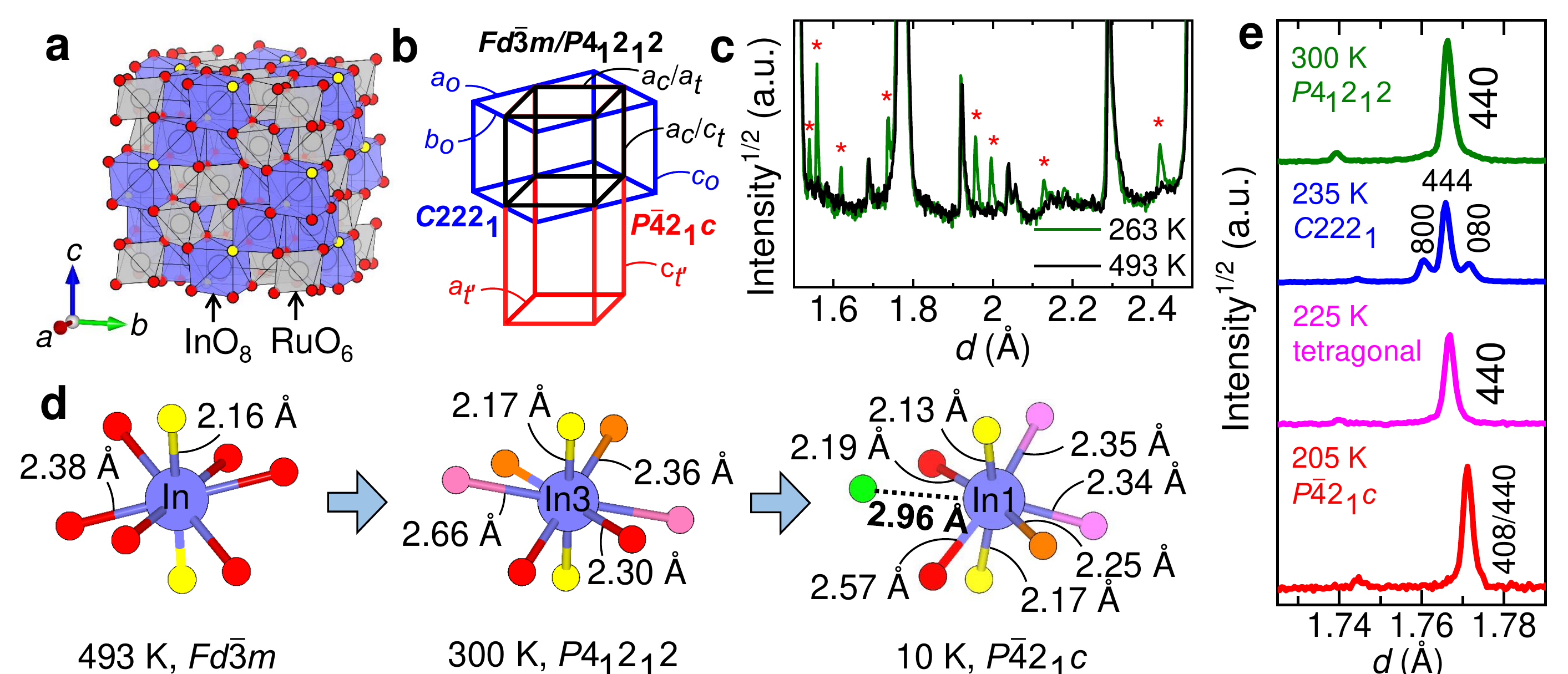}
	\caption{\textbf{Structural changes of In$_2$Ru$_2$O$_7$ through multiple phase transitions.} \textbf{a} Crystal structure of In$_2$Ru$_2$O$_7$ in the high-temperature cubic phase ($Fd\overline{3}m$). The red and yellow spheres represent two distinct oxygen sites, O1 and O2, respectively. \textbf{b} The relationship between unit cells for the cubic ($Fd\overline{3}m$, $a_c$), orthorhombic ($C222_1$, $a_o$, $b_o$ and $c_o$)  and two tetragonal structures ($P4_12_12$, $a(c)_t$ and $P\overline{4}2_1c$, $a(c)_{t'}$). $a_o, b_o \sim \sqrt{2}a_t$ and $c_{t'} \sim 2 c_t$. The tetragonal phase around room temperature ($P4_12_12$) has a unit cell metric nearly identical to that of the cubic phase. \textbf{c} powder X-ray diffraction patterns of In$_2$Ru$_2$O$_7$ in the cubic ($\SI{493}{\kelvin}$) and $P4_12_12$ tetragonal ($\SI{263}{\kelvin}$) phases. The peaks marked by asterisks are visible only in the $P4_12_12$ tetragonal phase. \textbf{d} The change of In-O bonding geometry with decreasing temperature. The yellow and other-coloured spheres denote the O2 or O2-derived and O1 or O1-derived oxygen atoms, respectively. In the $P\overline{4}2_1c$ phase at 10 K, the green oxygen atom (O4) is part of the short Ru3-O4-Ru3 bond shown in Fig.~\ref{fig:molecule}\textbf{a}. For more detail, see Supplementary Tables S3 and S4. \textbf{e} Neutron powder diffraction patterns of In$_2$Ru$_2$O$_7$ showing the evolution of the 440 reflection with decreasing temperature. The space group of the tetragonal phase at $\SI{225}{\kelvin}$ has not been identified yet.}
	\label{fig:stru}
\end{figure*}

To explore the novel phase competition in pyrochlore ruthenates and investigate the impact of $A$-O bond covalency, we synthesized a new compound In$_2$Ru$_2$O$_7$. The strong covalency of In-O bonds has been identified in the sister compound, a pyrochlore iridate In$_2$Ir$_2$O$_7$~\cite{Krajewska2020}. At high temperatures above $\SI{450}{\kelvin}$, In$_2$Ru$_2$O$_7$ crystallizes in a cubic pyrochlore structure, but adopts a distorted pyrochlore structure at room temperature. From spectroscopic measurements, In$_2$Ru$_2$O$_7$ was found to host a $J_{\rm eff}$ = 0-derived singlet state at room temperature despite the largest trigonal distortion among the family of pyrochlore ruthenates. The spin-orbit-entangled singlet state is not stable in In$_2$Ru$_2$O$_7$ at low temperatures. Through the successive structural transitions likely associated with the covalent In-O bonds, In$_2$Ru$_2$O$_7$ displays a nonmagnetic insulating state below $\SI{220}{\kelvin}$. The nonmagnetic ground state was found to originate from a molecular orbital formation within the semi-isolated Ru$_2$O units on the pyrochlore lattice. We argue that the unique molecular orbital formation involving oxygen atoms, distinct from the dimers with direct overlap of $d$-orbitals, is associated with the distortion of In-O network induced by bond covalency. Our result demonstrates that the bond character of constituent ions, which is often overlooked in the physics of complex transition-metal oxides, can be an important factor in electronic phase competition in those materials.

\begin{figure}[tb]
	\centering
	\includegraphics[width=1\linewidth]{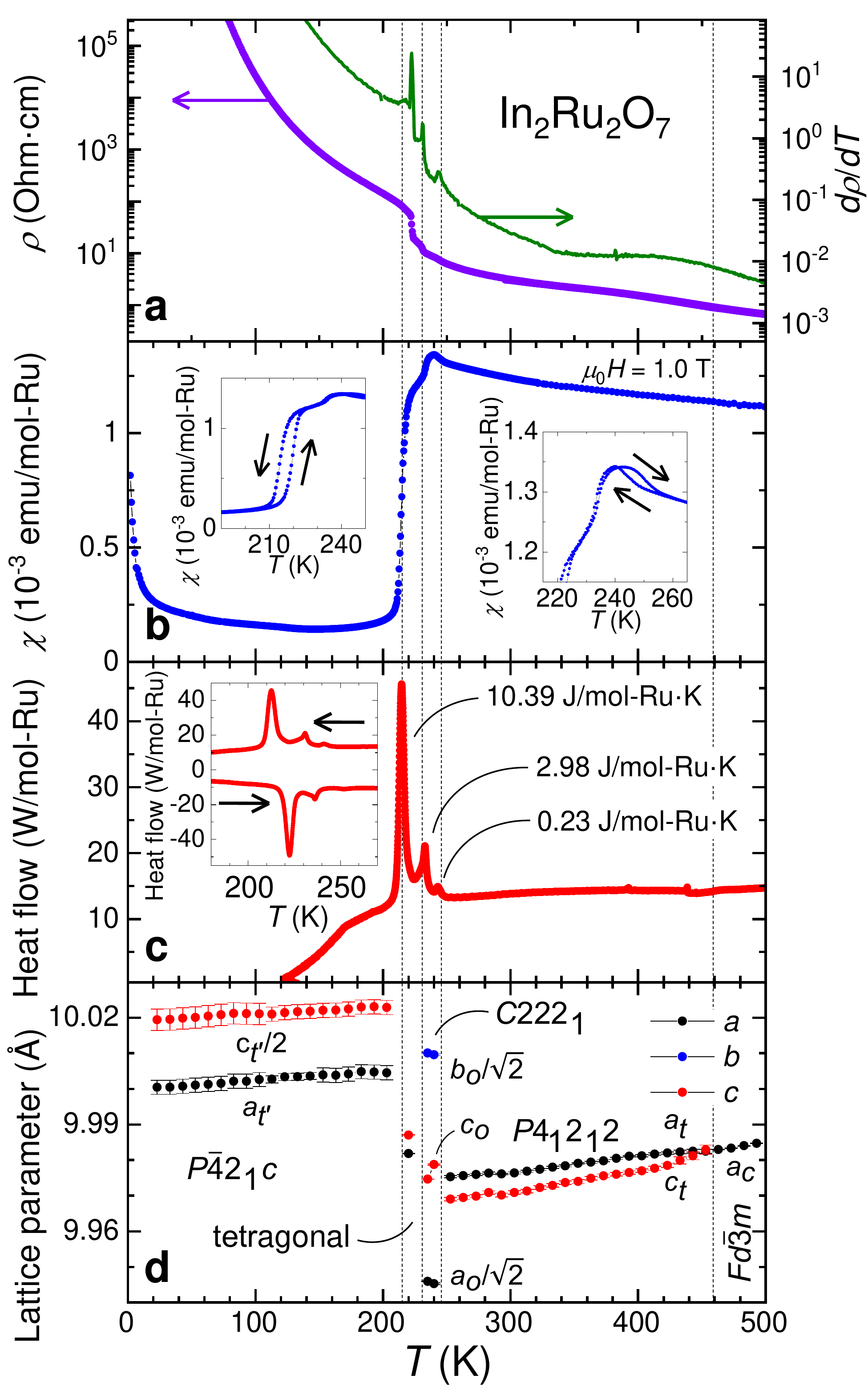}
	\caption{\textbf{Multiple phase transitions in In$_2$Ru$_2$O$_7$.}  Temperature-dependent  \textbf{a} resistivity $\rho(T)$ and the derivative of the resistivity $d\rho/dT$, \textbf{b} magnetic susceptibility $\chi(T)$, \textbf{c} heat flow determined with differential scanning calorimetry (DSC), and \textbf{d} the lattice parameters of polycrystalline In$_2$Ru$_2$O$_7$, respectively. The data in the main panel were collected in the cooling process. The insets of \textbf{b} show the thermal hysteresis of two transitions, where the black arrows denote the zero-field-cooling ($\rightarrow$) and field-cooling ($\leftarrow$) curves. In \textbf{c}, the entropy release identified in the three separate first order phase transitions is indicated, and the inset shows both the heating ($\rightarrow$) and cooling ($\leftarrow$) DSC curves. In \textbf{d}, the $a$, $b$ and $c$ lattice parameters of the tetragonal and orthorhombic unit cells are normalised with respect to those of the $Fd\overline{3}m$ cubic structure as indicated in the figure.}
	\label{fig:phys}
\end{figure}

\section{Results}
 
\subsection{Spin-orbit-entangled singlet state near room temperature}

The powder and single crystals of In$_2$Ru$_2$O$_7$ were synthesised by a solid-state reaction and a flux method under high-pressures of 6 and 8 GPa, respectively (see \textbf{Methods}). At room temperature, the powder X-ray diffraction pattern of the product indicated the pyrochlore-like structure but there are several small peaks which cannot be indexed by the cubic pyrochlore structure with the space group $Fd\overline{3}m$ (marked with red asterisks in Fig.~\ref{fig:stru}\textbf{c}). By increasing the temperature above $\SI{450}{\kelvin}$, the small peaks disappear and the Rietveld analysis of the X-ray and neutron powder diffraction patterns shows that the product is single phase cubic pyrochlore In$_2$Ru$_2$O$_7$ (Fig.~\ref{fig:stru}\textbf{a}) without any impurities within the instrumental resolution (see Supplementary Fig. S1 and Table S1). In$_2$Ru$_2$O$_7$ has the smallest cubic lattice parameter $a = \SI{9.9847(1)}{\angstrom}$ at $T = \SI{493}{\kelvin}$, the most trigonally compressed RuO$_6$ octahedra and the smallest Ru-O-Ru angle of $\SI{125.6(3)}{\degree}$ among the reported pyrochlore ruthenates. 

With decreasing temperature, In$_2$Ru$_2$O$_7$ was found to undergo a weak tetragonal distortion at $\sim\SI{450}{\kelvin}$ and the single crystal X-ray diffraction indicated a non-centrosymmetric structure with space group $P4_12_12$ (Supplementary Fig. S2 and Table S1). The unit cell remains approximately the same size (Fig.~\ref{fig:stru}\textbf{b}) and both In and Ru sites, which are originally 16-fold single sites in the cubic phase, split into two 4-fold and one 8-fold sites each. The changes in the Ru-O bond lengths and Ru-O-Ru bond angles are not appreciable, and the degree of Ru-O bond modulation is of the order of $\SI{3}{\percent}$ (Supplementary Fig. S3). In contrast, the coordination environment around the In atoms distorts significantly. In the cubic phase above $\SI{450}{\kelvin}$ there are two crystallographically inequivalent oxygen sites, O1 and O2. The Ru atoms are octahedrally coordinated by the O1 atoms, whereas the In atoms are coordinated by six O1 and two O2 atoms, forming a scalenohedron (Fig.~\ref{fig:stru}\textbf{a}). The In-O1 bonds of $\SI{2.375}{\angstrom}$ in the cubic phase disproportionate into shorter ($\SI{2.30}{\angstrom}$) and longer ($\SI{2.66}{\angstrom}$) bonds with a difference of as much as $\SI{10}{\percent}$ in the $P4_12_12$ structure (Fig.~\ref{fig:stru}\textbf{d}).  

\begin{figure*}[t]
	\centering
	\includegraphics[width=1\linewidth,trim= 0 120 130 0,clip]{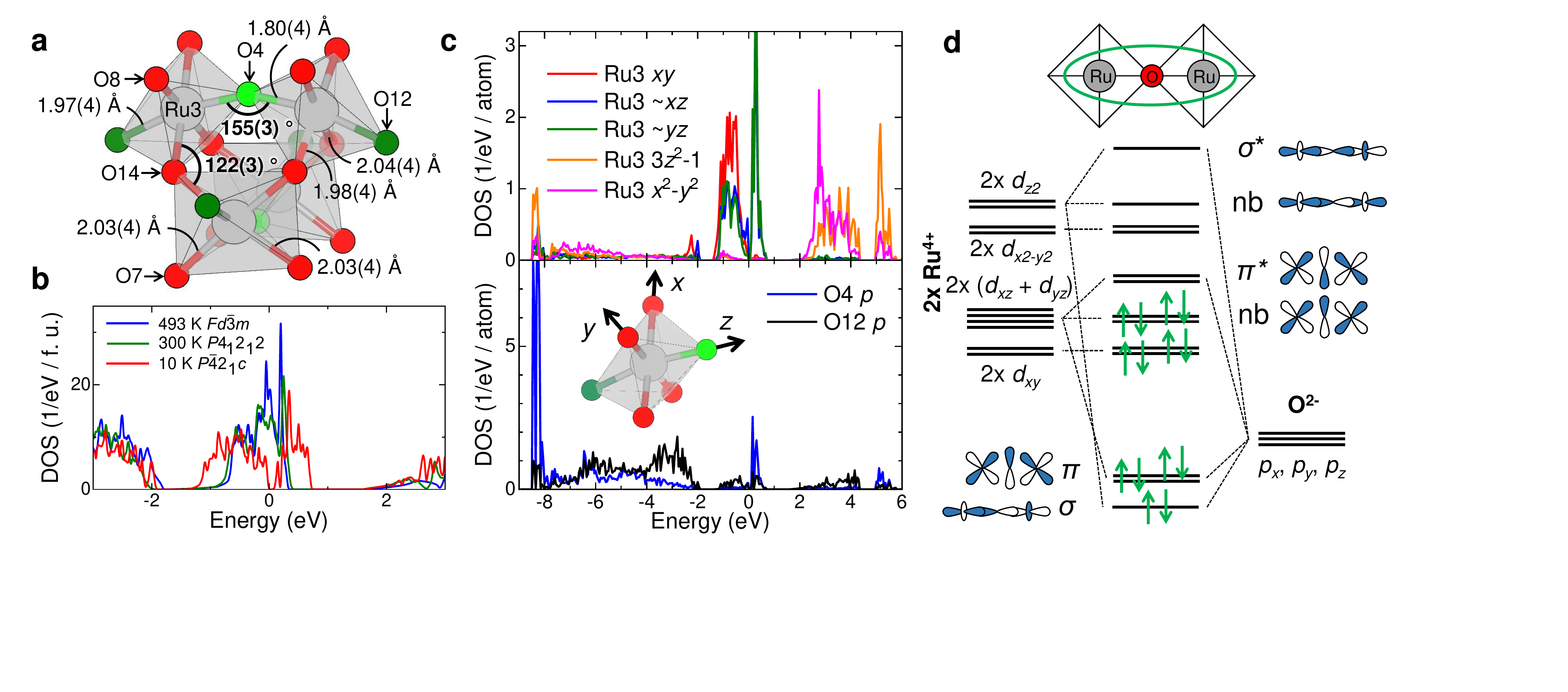}
	\caption{\textbf{Formation of Ru$_2$O orbital molecules on the pyrochlore lattice.} \textbf{a} Crystal structure in the vicinity of the Ru3-O4-Ru3 bond in the low-temperature tetragonal phase of In$_2$Ru$_2$O$_7$. Four Ru3O$_6$ octahedra, where the four Ru3 atoms form a tetrahedron within the pyrochlore lattice, are shown. \textbf{b} The total density of states per formula unit (f.u.) of the high-temperature cubic ($Fd\overline{3}m$), room-temperature tetragonal ($P4_12_12$), and low-temperature tetragonal ($P\overline{4}2_1c$) phases of In$_2$Ru$_2$O$_7$. \textbf{c} Density of states projected onto Ru3 $4d$ states (upper) and O 2$p$ states of O4 and O12 coordinating Ru3 (lower), obtained by a scalar-relativistic calculation. The $x$, $y$ and $z$ axes point along the Ru-O bonds with the $z$ axis parallel to the Ru-O4 bond, as illustrated in the inset. \textbf{d} Molecular orbital (MO) scheme of Ru$_2$O unit with $\SI{180}{\degree}$ Ru-O-Ru geometry. $\pi$-type Ru $t_{2g}$ - O $p$ and $\sigma$-type  Ru $e_{g}$ - O $p$ hybridisation is considered. $\sigma^{\ast}$ and $\pi^{\ast}$ denote antibonding MO, while nb corresponds to non-bonding MO. The individual Ru $d$ levels are split due to pseudo-tetragonal distortion prior to Ru-O hybridisation. As each Ru$^{4+}$ ion provides 4 $d$-electrons and an O$^{2-}$ ion contributes 6 $p$-electrons, there are 14 electrons in total per Ru$_2$O molecule.}
	\label{fig:molecule}
\end{figure*}

Above room temperature, the resistivity $\rho(T)$ of In$_2$Ru$_2$O$_7$ shows an insulating behaviour with an activation gap of $\sim$150 meV, and the magnetic susceptibility $\chi(T)$ shows a Curie-Weiss-like increase on cooling as shown in Fig.~\ref{fig:phys}\textbf{a} and \textbf{b}. The $Fd\overline{3}m \rightarrow P4_12_12$ structural transition at $\sim \SI{450}{\kelvin}$ does not exhibit pronounced anomalies in $\rho(T)$ and $\chi(T)$, and is of second order as supported by the absence of a visible peak in differential scanning calorimetry (DSC) at this temperature (Fig.~\ref{fig:phys}\textbf{c}). The density functional theory calculations for the $Fd\overline{3}m$ and $P4_12_12$ phases without Coulomb repulsion $U$ indicate a metallic state with a sizeable density of states at the Fermi energy $E_{\rm F}$ (Fig.~\ref{fig:molecule}\textbf{b}), pointing to a Mott insulating state in these phases. In addition, the calculated band structures of the two phases show no significant differences near $E_{\rm F}$ (Supplementary Fig. S4), implying that the phase transition is not electronic in origin and is caused by the lattice instability of the In-O bonding geometry.

\begin{figure*}[t]
	\centering
	\includegraphics[width=0.75\linewidth,trim=0 0 510 0,clip]{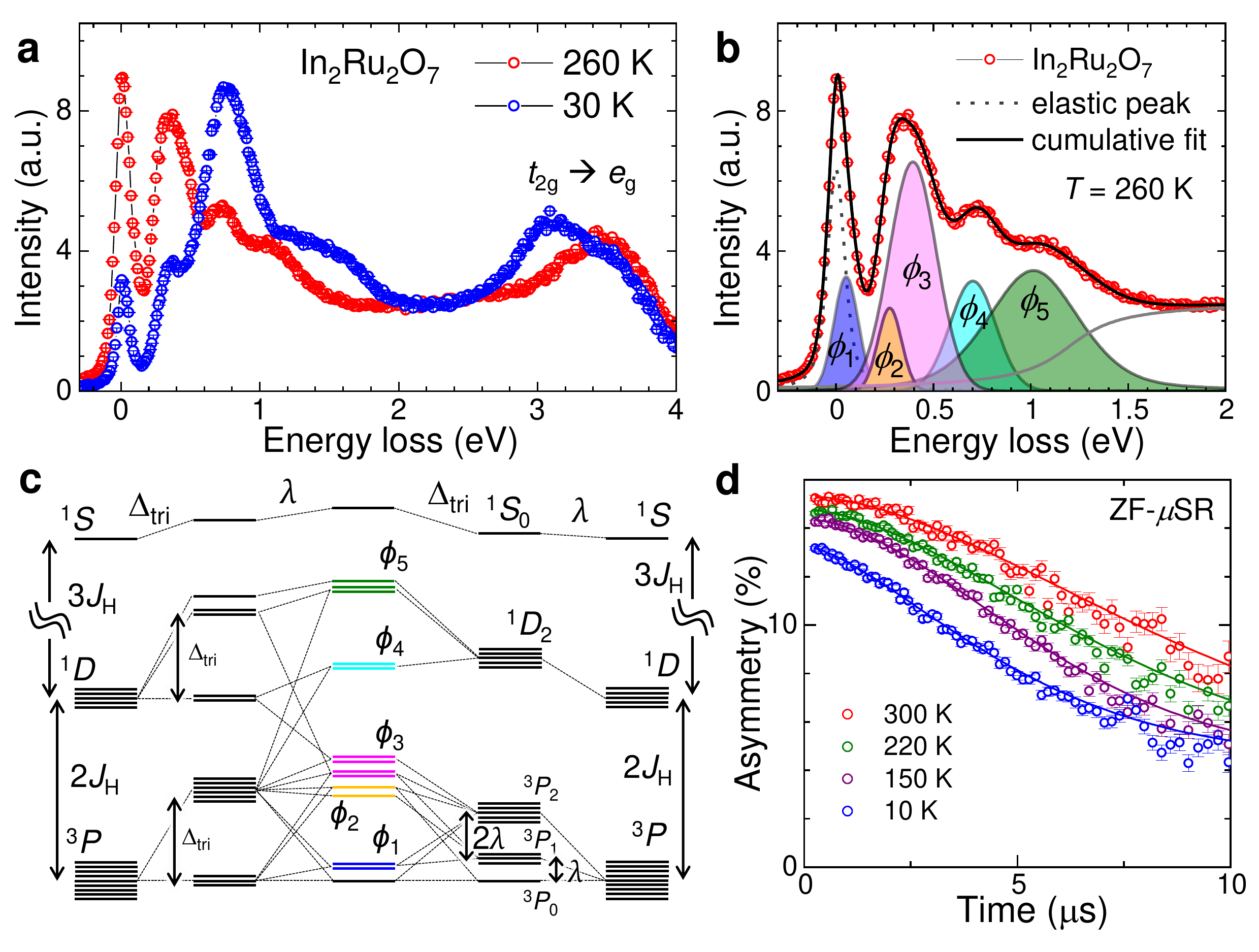}
	\caption{\textbf{Electronic structure of In$_2$Ru$_2$O$_7$.} \textbf{a} Ru $L_3$-edge RIXS spectra of polycrystalline In$_2$Ru$_2$O$_7$ with incident energy $E_i = \SI{2839.2}{eV}$ at $\SI{260}{\kelvin}$ and $\SI{30}{\kelvin}$. \textbf{b} Fitting of RIXS spectrum at $T = \SI{260}{\kelvin}$. The dashed line shows the fitted elastic peak, while the coloured peaks correspond to the excitations which are schematically shown in \textbf{c}. The grey line corresponds to an arctangent function representing the electronic continuum. All peaks were fitted with a pseudo-Voigt profile. \textbf{c} The energy level diagram of the $t_{2g}^{4}$ electronic configuration in an isolated RuO$_6$ octahedron ($J_H >> \lambda_{\mathrm{SOC}}$ and 10Dq = $\infty$) formed with two different procedures. Left: the $t_{2g}^4$ states are split by Hund's coupling $J_{\mathrm{H}}$, then by compressive trigonal crystal field $\Delta_{\mathrm{tri}}$ and finally by spin-orbit coupling $\lambda_{\mathrm{SOC}}$. Right: the $t_{2g}^4$ are split by $J_{\mathrm{H}}$, then by $\lambda_{\mathrm{SOC}}$ and finally by  $\Delta_{\mathrm{tri}}$. The dashed lines indicates the mixing of states. The $\phi_1$, $\phi_2$, $\phi_3$, $\phi_4$ and $\phi_5$ coloured states correspond to the labelled excitations in \textbf{b}. \textbf{d} ZF-$\mu$SR time spectrum of polycrystalline In$_2$Ru$_2$O$_7$ at various temperatures. The solid lines show a stretched exponential fit of the data in the form $P_z(t) = Ae^{-({\lambda{t}})^{\beta}}$, where $P_z(t)$ is the time-dependent asymmetry, $A$ is the initial asymmetry, $\lambda$ is the relaxation rate and $\beta$ is the exponent (see Supplementary Fig. S11 for the fitted values).}
	\label{fig:RIXS}
\end{figure*}

The Curie-Weiss fit of $\chi(T)$ at high temperatures ($> \SI{500}{\kelvin}$) yields an effective moment $\mu_{\mathrm{eff}}$ of \SI{3.83(2)}{\bohrmagneton} which exceeds that of $S = 1$, implying that the the fit is performed well below the Curie-Weiss regime, or that the magnetic properties cannot be accounted for by spin-only moments. In order to study the electronic structure of In$_2$Ru$_2$O$_7$, we performed resonant inelastic X-ray scattering (RIXS) measurement at the Ru $L_3$ edge (Fig.~\ref{fig:RIXS}\textbf{a}). At $T = \SI{260}{\kelvin}$, namely in the $P4_12_12$ phase, we resolved six excitation peaks centred at \SI{53(17)}{\meV}, \SI{276(40)}{\meV}, \SI{394(34)}{\meV}, \SI{700(18)}{\meV}, \SI{1.01(2)}{\eV} and $\sim \SI{3.5}{\eV}$ in addition to the elastic one. While the broad peak at $\SI{3.5}{\eV}$ represents the excitations from the $t_{2g}$ to $e_{g}$ manifold, namely 10Dq, the ones below $\SI{1.5}{\eV}$ correspond to excitations within the $t_{2g}$ manifold. The $t_{2g}$ multiplet excitations are labeled $\phi_1$, $\phi_2$, $\phi_3$, $\phi_4$ and $\phi_5$ as shown in Fig.~\ref{fig:RIXS}\textbf{b}. Our band structure calculation of the $Fd\overline{3}m$ phase without spin-orbit coupling shows a split of the $t_{2g}$ manifold into the $a_{1g}$ and $e_{g}$ states by the trigonal crystal field and the energy difference between the two manifolds is $\sim \SI{200}{\meV}$. This trigonal crystal field alone does not explain the lowest excitation at $\SI{53}{\meV}$. Considering the similar band structures between the $Fd\overline{3}m$ and $P4_12_12$ phases, the impact of a weak tetragonal distortion on the electronic structure should not be appreciable. Most likely, the effect of spin-orbit coupling needs to be incorporated.

To describe the excitations observed below $\SI{1.5}{\eV}$, we employed a standard Hamiltonian for $d^4$ configuration with Hund's coupling $J_{\rm H}$, spin-orbit coupling $\lambda_{\mathrm{SOC}}$ and compressive trigonal field $\Delta_{\rm tri}$ (see \textbf{Methods}). We assume that 10Dq $>>J_{\rm H}$ and that all four electrons reside in the $t_{2g}$ manifold. First, in the $LS$ coupling limit, Hund's rules render the ${^{3}P}$ configuration ($S = 1, L_{\rm eff} = 1$) as the ground state, the ${^{1}D}$ configuration ($S = 0, L_{\rm eff} = 2$) as the next excited state and the ${^{1}S}$ configuration ($S = 0, L_{\rm eff} = 0$) as the highest energy state as depicted in Fig.~\ref{fig:RIXS}\textbf{c}. Second, the compressive trigonal crystal field splits the ${^{3}P}$ and ${^{1}D}$ multiplets into a triplet and a sextet, and two doublets and a singlet, respectively. The lowest-lying triplet approximately corresponds to $L = 0$, $S = 1$ configuration. Finally, $\lambda_{\mathrm{SOC}}$ mixes these states and further splits the lowest triplet into a singlet and a doublet and then the sextet into two singlets and two doublets which are shown in the diagram at the centre of Fig.~\ref{fig:RIXS}\textbf{c}. The ground state is a spin-orbit-entangled singlet and a doublet (effective $S^z = \pm 1$) is the lowest excited state. Therefore, the excitations observed at $T = \SI{260}{\kelvin}$ can be labelled as follows: $\phi_1$ is the excitation to the lowest doublet and $\phi_2$ and $\phi_3$ correspond to the excitations to the states split from the (${^{3}P}$)-derived sextet. $\phi_4$ and $\phi_5$ correspond to the excitations to the states split from the (${^{1}D}$)-derived states. We note that the same energy diagram can be obtained when we introduce spin-orbit coupling right after the Hund's coupling and then add the trigonal crystal field, i.e. via the spliting of the $J_{\rm eff}$-states by the trigonal crystal field, as shown in the right side of Fig.~\ref{fig:RIXS}\textbf{c}.

By optimising the relevant parameters to the observed peak positions, we obtain the values of $J_{\rm H} \sim \SI{310}{\meV}$, $\lambda_{\mathrm{SOC}} \sim \SI{70}{\meV}$ and $\Delta_{\rm tri} \sim \SI{300}{\meV}$, similar to those reported in Ca$_2$RuO$_4$~\cite{Gretarsson2019}. The magnitude of the non-cubic crystal field is larger in In$_2$Ru$_2$O$_7$ than that reported in Ca$_2$RuO$_4$, which can be rationalised by the strong trigonal compression of the RuO$_6$ octahedra. The single-ion model including spin-orbit coupling and trigonal crystal field therefore explains the RIXS spectrum well, which demonstrates that the orbital degree of freedom is unquenched in In$_2$Ru$_2$O$_7$ and a spin-orbit-entangled singlet state is realised at room temperature.

The behaviour of $\chi(T)$ above room temperature can be understood as van Vleck susceptibility arising from the excitation between the ground state singlet and the lowest doublet. Considering the presence of long-range magnetic order in other pyrochlore ruthenates~\cite{Sato1999,Taira2002}, the magnetic exchange coupling may be strong enough compared to the magnetic excitation gap. Therefore, the van Vleck moments residing on the singlet-doublet transition can be induced not only by an external magnetic field, but also by exchange interactions as well~\cite{Khaliullin2013,Chaloupka2019}. In addition, the small gap of $\sim \SI{50}{\meV}$ implies that there would be sizeable contributions of magnetic moments from the thermally-excited doublets at high temperatures.

\subsection{Multiple phase transitions in In$_2$Ru$_2$O$_7$ below room temperature}

A spin-orbit-entangled singlet state is expected to display excitonic magnetism at low temperatures via the condensation of excited states. On cooling from $\SI{300}{\kelvin}$, In$_2$Ru$_2$O$_7$ undergoes multiple phase transitions. First, at around $\SI{250}{\kelvin}$ a small hump with hysteresis in $\chi(T)$ (the right inset of Fig.~\ref{fig:phys}\textbf{b}) and a kink in $\rho(T)$ are observed (Fig.~\ref{fig:phys}\textbf{a}). Simultaneously, the crystal structure becomes orthorhombic (Space group $C222_1$), adopting $a_o\sim\sqrt{2}a_t$, $b_o\sim\sqrt{2}a_t$, $c_o \sim c_t$ unit cell as evident from the splitting of the reflections in the neutron diffraction pattern at $\SI{235}{\kelvin}$ (Fig.~\ref{fig:stru}\textbf{e}). Second, at about $\SI{230}{\kelvin}$, $\chi(T)$ shows a moderate drop, while $\rho(T)$ exhibits a small jump. At that temperature, the diffraction peaks associated with the orthorhombic distortion vanish, implying that In$_2$Ru$_2$O$_7$ adopts a primitive tetragonal cell. As evident from DSC, a total entropy of $\sim$3 J/mol-Ru$\cdot$K is released when crossing these two transitions, which could suggest partial quenching of orbital and/or spin degrees of freedom. 

Upon further cooling, a sharp drop of $\chi(T)$ down to a value of $\sim$ 2.5$\times$10$^{-4}$ emu/mol-Ru is observed as shown in Fig.~\ref{fig:phys}\textbf{b}. Below $\SI{220}{\kelvin}$, $\chi(T)$ exhibits nearly no temperature dependence except for the low-temperature upturn likely due to defects or minor impurities. No further anomalies are seen in specific heat $C(T)$ below $\SI{220}{\kelvin}$ (Supplementary Fig. S5). This suggests the emergence of a nonmagnetic state or long-range antiferromagnetic order at the transition. $\rho(T)$ shows a discontinuous jump associated with the transition, and the thermal hysteresis of $\chi(T)$ (the left inset of Fig.~\ref{fig:phys}\textbf{b}) and a large peak in DSC with an entropy release of as large as 10.39 J/mol-Ru$\cdot$K indicate that the transition is first order and involves the quenching of spin and orbital degrees of freedom. We note that on cooling from high temperature across the multiple transitions, a total of $\sim$13.5 J/mol-Ru$\cdot$K is released, which is close to the total entropy of $^3P$ spin-orbital manifold ($L_{\mathrm{eff}}=1$, $S=1$), namely $R$ln9 of the $\sim$ 18 J/mol-Ru$\cdot$K. 

To identify the nature of the low-temperature phase, we performed muon spin rotation measurements at zero magnetic field (ZF-$\mu$SR) on a polycrystalline In$_2$Ru$_2$O$_7$. The time dependence of the muon asymmetry did not show an oscillation down to $\SI{10}{\kelvin}$ (Fig.~\ref{fig:RIXS}\textbf{d}), which excludes the presence of long-range magnetic order and points to a nonmagnetic state. In addition, the RIXS spectrum measured at $\SI{30}{\kelvin}$ displays drastic changes (Fig.~\ref{fig:RIXS}\textbf{a}). A large portion of the spectral weight appears to be transferred to higher energies, suggesting that the local spin-orbit-entangled singlet is no longer valid.

\subsection{Formation of the Ru$_2$O orbital molecules at low temperatures}

To unveil the nature of the low-temperature nonmagnetic state, we investigated the crystal structure below $\SI{220}{\kelvin}$. The single crystal X-ray diffraction, together with powder neutron diffraction, showed that the crystal symmetry of In$_2$Ru$_2$O$_7$ is further reduced to the space group $P\overline{4}2_1c$ with the doubling of the unit cell (Fig.~\ref{fig:stru}\textbf{b,e} and Supplementary Fig. S2). The $P\overline{4}2_1c$ unit cell adopts lattice parameters of $a = \SI{10.00415(17)}{\angstrom}$ and $c = \SI{20.0148(5)}{\angstrom}$, as refined at $\SI{10}{\kelvin}$, and hosts four inequivalent 8-fold In and Ru crystallographic sites (Supplementary Table S2). For simplicity, we firstly discuss the local structure around one of the Ru sites, Ru3. As shown in Fig.~\ref{fig:molecule}\textbf{a}, we find that (Ru3)O$_6$ octahedron undergoes a pseudo-tetragonal compression, where two ``apical'' oxygen atoms, O4 and O12 (light green and dark green, respectively), move closer to the Ru atom, and four other oxygen atoms, O7, O8 and two O14 (red), move away from Ru. In particular, the axial Ru3-O4 bond becomes shorter by about $\SI{10}{\percent}$ as compared to the other five Ru3-O bonds. The O4 atoms involved in the short bond form an identical Ru3-O4 bond in the neighbouring (Ru3)O$_6$ octahedron, giving a short Ru3-O4-Ru3 link. In the high-temperature cubic structure, this oxygen atom coordinates two Ru atoms and two In atoms simultaneously. In the $P\overline{4}2_1c$ structure, it is displaced much closer towards the Ru atoms, and effectively no longer coordinates to the In atoms, where the In-O4 distance increases up to $\sim$3 ${\rm \AA}$ (Fig.~\ref{fig:stru}\textbf{d}). As a result of this displacement, the Ru3-O4-Ru3 angle becomes as large as $\sim \SI{155}{\degree}$, while other Ru-O-Ru angles are $\sim \SI{125}{\degree}$, close to the value of the cubic phase. We find that while there are four inequivalent Ru sites, all of them participate in such short and large angle Ru-O-Ru bonding geometry such as Ru2-O6-Ru2 and Ru1-O11-Ru4 bonds (see Supplementary Fig. S6). Although the details of bond distance and angles are different at each Ru site, the observed structural features are qualitatively the same. In contrast to the significant change of the Ru-O bonds, the Ru-Ru distances do not change appreciably throughout the structural transitions as corroborated by the analysis of X-ray absorption fine structure data (Supplementary Fig. S7). We call these Ru$_2$O units ``Ru$_2$O molecules'' based on the formation of molecular orbitals as described below. The Ru$_2$O molecules are effectively isolated from each other because of much larger inter-molecule Ru-O bond lengths. Consequently, the Ru-O pyrochlore lattice of In$_2$Ru$_2$O$_7$ transforms into an arrangement of semi-isolated Ru$_2$O molecules. 

The pronounced distortion of the Ru-O network at low temperatures has a strong impact on the electronic structure. The calculated band structures for the high-temperature and room-temperature phases ($Fd\overline{3}m$ and $P4_12_12$ structures) are metallic when on-site Coulomb energy $U$ is not incorporated (Fig.~\ref{fig:molecule}\textbf{b}). In contrast, the low-temperature structure comprising Ru$_2$O molecules gives an insulating ground state even without the inclusion of $U$. The DOS projected onto the Ru $4d$ orbitals and the O $2p$ orbitals for one RuO$_6$ octaheron of the low-temperature phase, Ru3O$_6$, is shown in Fig.~\ref{fig:molecule}\textbf{c}. The partial DOS of the Ru 4$d$ states are resolved into $d_{xy}$, $d_{xz}$, $d_{yz}$, $d_{z^2}$ and $d_{x^2 - y^2}$ orbitals, where the $z$ axis points along the shortest Ru3-O4 bond. For the $t_{2g}$ orbitals, the $d_{xy}$ orbital has a lower energy due to the pseudo-tetragonal compression of the RuO$_6$ octahedron and is fully occupied, while the $d_{xz}$ and $d_{yz}$ orbitals are split, yielding a small charge gap. The energy difference between the centroids of occupied and unoccupied $t_{2g}$ manifolds is about $\SI{1}{\eV}$ (Supplementary Fig. S8), which is close to the most pronounced excitation at $\sim \SI{0.75}{\eV}$ in the RIXS spectrum at $T = \SI{30}{\kelvin}$ (Fig.~\ref{fig:RIXS}\textbf{a}). 

The presence of a charge gap without $U$ can be attributed to the formation of molecular orbitals (MO) within the Ru$_2$O units. For simplicity, we consider a Ru$_2$O molecule with $\SI{180}{\degree}$ bond geometry embedded in two corner-sharing RuO$_6$ octahedra, where the Ru $d$-orbitals are split into $t_{2g}$ and $e_g$ manifold and the tetragonal distortion of RuO$_6$ is incorporated. The expected MO levels are depicted in Fig.~\ref{fig:molecule}\textbf{d}. One of the $d_{z^2}$ orbitals forms a strong $\sigma$-type bonding and anti-bonding orbitals with the O $2p_z$ orbital, while the other one remains non-bonding because of an out-of-phase configuration as in a linear H$_3$ molecule. In contrast, the $d_{x^2-y^2}$ orbital does not hybridise with the O 2$p$ orbitals and thus remains non-bonding. For the $t_{2g}$ orbitals, the $d_{xy}$ orbitals form non-bonding orbitals without any hybridisation with the O 2$p$ orbitals. The $d_{zx}$ and $d_{yz}$ orbitals hybridise with the O 2$p_x$ and 2$p_y$ states, respectively, and give rise to bonding, non-bonding and antibonding $\pi$-type MOs, where the latter two are largely of $d$-character. As there are 14 electrons in a Ru$_2$O molecule, four 4$d$-electrons from each Ru$^{4+}$ ion and six 2$p$-electrons from an O$^{2-}$ ion, those electrons fill the MOs up to the non-bonding $d_{xz}$($d_{yz}$) orbitals and a gap is formed between the non-bonding and the anti-bonding $\pi$ orbitals.

Although the Ru-O-Ru angle within the actual Ru$_2$O molecules deviates from $\SI{180}{\degree}$, these MO characters are indeed identified in the calculated DOS of the $P\overline{4}2_1c$ structure (Fig.~\ref{fig:molecule}\textbf{c}). The $\sigma$ and $\pi$ bonding orbitals, which have strong O 2$p$ contributions, can be seen in the partial DOS of the O4 atom incorporated in the Ru$_2$O molecule as a sharp peak around $\SI{-8}{\eV}$ and a broad feature from $\SI{-8}{\eV}$ to $\SI{-2}{\eV}$, respectively (lower panel). The non-bonding $d_{z^2}$ and $d_{x^2-y^2}$ orbitals and the $\sigma$ anti-bonding orbital primarily of $d_{z^2}$ character are present in the partial DOS of the Ru3 4$d$ orbitals at 2-$\SI{4}{\eV}$ and 5-$\SI{6}{\eV}$ above $E_{\rm F}$, respectively. The splitting of two non-bonding $e_{g}$ states around 2-$\SI{4}{\eV}$ is due to the pseudo-tetragonal distortion, which results in the split of the $t_{2g}$$\rightarrow$$e_{g}$ excitation in the RIXS spectrum at low temperature (Fig.~\ref{fig:RIXS}\textbf{a}). In the vicinity of $E_{\rm F}$, the $d_{xy}$ states and occupied $d_{yz}$/$d_{zx}$ states correspond to the non-bonding $t_{2g}$ orbitals in the level scheme of Fig.~\ref{fig:molecule}\textbf{d}, which is corroborated by the negligible contribution of the O4 2$p$ states in the energy region of -1 to $\SI{0}{\eV}$. On the other hand, one can identify a sizeable contribution of O4 2$p$ states right above $E_{\rm F}$ where unoccupied $d_{yz}$/$d_{zx}$ states reside, indicating that the states right above $E_{\rm F}$ are derived from the anti-bonding states between Ru3 4$d_{yz}$($d_{zx}$) and O4 2$p_y$($p_x$) orbitals. Therefore, although the details of the electronic structure may be altered by the presence of modest $U$ and spin-orbit coupling $\lambda_{\rm SO}$, the origin of the nonmagnetic insulating ground state of In$_2$Ru$_2$O$_7$ can be attributed to the formation of molecular orbitals in the Ru$_2$O units, namely Ru$_2$O orbital molecules, produced by the strong structural distortion.

\section{Discussion}

The new pyrochlore ruthenate In$_2$Ru$_2$O$_7$ displays a competition between a spin-orbit-entangled singlet state and a molecular orbital formation. Despite the largest trigonal distortion among the family of pyrochlore ruthenates, spin-orbit coupling $\lambda_{\rm SOC}$ is still operative and produces a spin-orbit-entangled singlet state at room temperature. This result suggests that other pyrochlore ruthenates $A_2$Ru$_2$O$_7$ with reduced trigonal distortion also have a spin-orbit-entangled singlet state at high temperatures and their long-range magnetic ordering may represent excitonic magnetism (see Supplementary Fig. S9 for the electronic structure of Y$_2$Ru$_2$O$_7$). In contrast to those pyrochlore ruthenates, In$_2$Ru$_2$O$_7$ displays multiple structural transitions and eventually becomes a nonmagnetic insulator accompanied by formation of Ru$_2$O orbital molecules. The origin of two transitions at $\SI{235}{\kelvin}$ and $\SI{225}{\kelvin}$ is not clear yet, and further structural analysis is required to unravel the nature of the intermediate phases. As we observe substantial entropy release even above $\SI{220}{\kelvin}$, we expect that the two transitions involve quenching of spin and orbital degrees of freedom, likely being a precursor for the Ru$_2$O orbital molecules. 

The competition between spin-orbital entanglement and molecular orbital formation has been seen in a number of 4$d$ or 5$d$-based systems such as honeycomb-based iridates and ruthenates ~\cite{Hermann2019,Bastien2018, Miura2007,Miura2009,Takayama2019,Veiga2019}. A distinct feature in In$_2$Ru$_2$O$_7$ is that the molecular orbitals are formed with the involvement of the oxygen 2$p$ electrons, namely induced by Ru 4$d$-O 2$p$-Ru 4$d$ hopping through the corner-sharing bond, whereas the direct $d$-$d$ hopping across the shared edges of octahedra plays a crucial role in the honeycomb-based systems. 

To the best of our knowledege, this type of molecular orbital formation has never been observed in other pyrochlore ruthenates or other pyrochlore transition-metal oxides. We argue that this unique feature of In$_2$Ru$_2$O$_7$ can be attributed to the nature of In-O bonds. When the $A$-site cation is a rare-earth ion or Y$^{3+}$, the $A$-O bond is ionic and the bonding geometry remains intact down to low temperatures. By contrast, the structural transitions of In$_2$Ru$_2$O$_7$ are associated with the pronounced distortion of the In-O environment accompanied by the In-O bond disproportionation. Since the In 5$s$ orbital locates close to the O 2$p$ level (see Supplementary Fig. S10 for the orbital-resolved DOS including In 5$s$ and 5$p$ states), the In-O bonds are expected to have a stronger covalent character compared to $A$-O bonds in other pyrochlore oxides~\cite{Krajewska2020, Fujita2017}. Therefore the disproportionation of In-O bonds may be interpreted as the formation of strong covalent bonds with short bond-lengths at the expense of a decrease of the coordination number to avoid the repulsion between oxygen atoms.

In the low-temperature $P\overline{4}2_1c$ phase below $\SI{220}{\kelvin}$, the disproportionation of In-O bonds is further enhanced (Fig.~\ref{fig:stru}\textbf{d}). One oxygen atom that is involved in the Ru$_2$O orbital molecule is displaced substantially away from the In atom ($\sim \SI{3}{\angstrom}$) and essentially is no longer bonded to the In atoms. Meanwhile, the two In-O bonds which are derived from the longer In-O1 bonds of the cubic phase become short ($\SI{2.19}{\angstrom}$ and $\SI{2.25}{\angstrom}$) and comparable to those derived from the short In-O2 bonds of the cubic phase ($\SI{2.16}{\angstrom}$). The oxygen atoms forming the former short bonds are located in the local $xy$-plane of the RuO$_6$ octahedra (red-coloured oxygen atoms in Fig.~\ref{fig:molecule}\textbf{a}). These in-plane oxygen atoms should form a strong covalent bond with the In atom, which in turn reduces the hybridisation with the Ru 4$d$ states. The suppressed $d$-$p$ hybridisation of the in-plane oxygen atoms enhances the isolation of the Ru$_2$O molecules from the other constituents of the crystalline lattice. Accordingly, the strong distortion of the In-O network, originating from the In-O bond covalency, results in the formation of isolated Ru$_2$O orbital molecules on the pyrochlore lattice. A similar structural distortion with disproportionation of In-O bonds has been discussed in nonmagnetic indate spinels~\cite{Goodenough1955}, which could share a common mechanism with the transitions in In$_2$Ru$_2$O$_7$.

In$_2$Ru$_2$O$_7$ therefore offers a unique playground where the competition between electronic phases is associated with In-O bond covalency. Among the pyrochlore ruthenates, Tl$_2$Ru$_2$O$_7$ displays a similar nonmagnetic singlet ground state below \SI{120}{\kelvin}~\cite{Takeda1998,Lee2001}, which has been discussed as originating from Haldane gap formation on the Ru pyrochlore lattice ~\cite{Lee2006}. As Tl is located right below In in the periodic table, a similarly strong covalent Tl-O bond should be expected but the role of Tl in the context of the nonmagnetic state has not been intensively discussed, and possibly is worthy of further investigation. The bonding character of $A$-O bonds is thus an interesting player in electronic phase competitions in pyrochlore oxides. Such a bonding character of ``$A$-site'' cations may also be an important factor in other complex transition-metal oxides and should be taken into account in the design of novel quantum materials.

\section{Methods}

\subsection{Sample preparation}
The polycrystalline and single crystal samples of In$_2$Ru$_2$O$_7$ were obtained by a high-pressure synthesis technique. For the polycrystalline sample, powders of In$_2$O$_3$ (Alfa Aesar, $\SI{99.994}{\percent}$ metal basis) and RuO$_2$ (Tanaka Kikinzoku, $\SI{76}{\wtpercent}$ Ru content) were mixed and ground in stoichiometric molar ratios and sealed in a Pt foil ampoule. The ampoule was heated at $\SI{1500}{\celsius}$ for 60 minutes under a pressure of $\SI{6}{\GPa}$. For the single crystal growth, powders of In$_2$O$_3$, RuO$_2$ and InCl$_3$ flux (Alfa Aesar,  $\SI{99.99}{\percent}$ metal basis) were mixed and ground in 2:2:1 molar ratio. The ampoule was then heated at $\SI{1500}{\celsius}$ for 60 minutes under pressure of $\SI{8}{\GPa}$ and the InCl$_3$ flux was removed with water after the synthesis. The polycrystalline sample of reference material Y$_2$Ru$_2$O$_7$ was prepared under the same conditions as In$_2$Ru$_2$O$_7$, the only difference being the Y$_2$O$_3$ (Alfa Aesar, $\SI{99.994}{\percent}$ metal basis) powder precursor.

\subsection{Single crystal X-ray diffraction and structure solution}
Single crystal X-ray diffraction data were collected with SMART-APEX-II CCD X-ray diffractometer (Bruker AXS, Karlsruhe, Germany) equipped with a N-Helix low-temperature device (Oxford Cryosystems) and SMART-APEX-I CCD X-ray diffractometer (Bruker AXS, Karlsruhe, Germany) equipped with Cryostream 700${}^{\mathrm{Plus}}$ cooling/heating device (Oxford Cryosystems). The reflection intensities were integrated with the \emph{SAINT} subprogram in the Bruker Suite software package. Multi-scan absorption corrections were applied using either \emph{SADABS} or \emph{TWINABS}. Structure refinement was performed by full-matrix least-squares fitting with the \emph{SHELXL} software package~\cite{Sheldrick2008,Sheldrick2015} and with JANA2006~\cite{Petricek2014}.

\subsection{Powder X-ray diffraction}
Powder X-ray diffraction data of In$_2$Ru$_2$O$_7$ were collected in Debye-Scherrer (DS) geometry with Stoe Stadi-P transmission diffractometer with primary beam Johann-type Ge(111) monochromator for Mo-${K}\alpha_{1}$-radiation and Mythen-1K PSD, equipped with a hot and cold air blower (Oxford Cryostream 800, Oxford Cryosystems).

\subsection{Neutron powder diffraction}
Neutron powder diffraction data of the powdered In$_2$Ru$_2$O$_7$ sample were collected at the HRPD beamline of the ISIS neutron and muon source. The time-of-flight data were collected over the 300-\SI{10}{\kelvin} temperature range in a flat-plate geometry and normalised and corrected using in-house software~\cite{Arnold2014}. Some regions containing peaks from the sample environment, such as vanadium, aluminium and stainless steel, were removed from the data prior to analysis. Rietveld analysis of the neutron diffraction data was performed with the GSAS program~\cite{Toby2001}.

\subsection{Physical property measurements}
Magnetization data were collected using a commercial magnetometer (MPMS, Quantum Design), resistivity and specific heat (thermal relaxation method) were measured using the Physical Property Measurement System (PPMS, Quantum Design). Resistivity measurements at temperatures above $\SI{350}{\kelvin}$ were performed using in-house equipment with a tube furnace. The high temperature resistivity data was scaled against the low temperature resistivity data obtained with the PPMS via overlapping temperature regions of 300-$\SI{350}{\kelvin}$. Differential scanning calorimetry was performed using a DSC1 (Mettler Toledo).

\subsection{Resonant inelastic X-ray scattering}
The temperature dependence of the electronic structure was investigated by measuring resonant inelastic X-ray scattering (RIXS) spectra at the Ru $L_3$-edge at the P01 beamline at the PETRA-III synchrotron at DESY. The energy of the incident X-rays was tuned to $\SI{2838.5}{\eV}$ ($2p_{3/2} \rightarrow 4d(t_{2g})$ excitation) by using a primary Si(111) two-bounce monochromator (cryogenically cooled) and a secondary four-crystal asymmetrical Si(111) channel-cut monochromator. The incident X-ray beam was then focused with a Kirkpatrick-Baez mirror. The scattering angle was fixed to $\SI{90}{\degree}$ in horizontal scattering geometry. The scattered X-ray beam was analysed with a diced and spherically bent SiO$_2$(10$\overline{2}$) analyser. For more detail, see Ref.~\cite{Gretarsson2020}. The total energy resolution, evaluated by the elastic scattering from GE varnish, was $\SI{100(7)}{\meV}$. The measurements were performed on the polycrystalline pellet of In$_2$Ru$_2$O$_7$. The spectrum of polycrystalline Y$_2$Ru$_2$O$_7$ ~\cite{Kennedy1995} was also collected for a comparison (see Supplementary Fig. S9).

The RIXS spectrum of room temperature In$_2$Ru$_2$O$_7$ was theoretically interpreted by employing a standard Hamiltonian for $t_{2g}^4$ configuration ($L_{\mathrm{eff}} = 1$) which includes an intra-ionic Coulomb interaction, spin-orbit coupling, and trigonal crystal field. The Coulomb interaction is accounted for with the parameters $U$, $U'$ and $J_{\mathrm{H}}$ (under the approximation $U'=U-2J_{\mathrm{H}}$) in a form of a Kanamori Hamiltonian~\cite{Georges2013}:
\begin{multline}
H_{\mathrm{C}}
= U \sum\limits_{m}n_{m\up}n_{m\dn}
+ U' \sum\limits_{m{\neq}m'}n_{m\up}n_{m'\dn} \\
+ (U'-J_{\mathrm{H}}) \sum\limits_{m<m',{\sigma}}n_{m{\sigma}}n_{m'{\sigma}} \\
- J_{\mathrm{H}} \sum\limits_{m{\neq}m'}d^{\dagger}_{m\up}d_{m\dn}d^{\dagger}_{m'\dn}d_{m'\up}
+ J_{\mathrm{H}} \sum\limits_{m{\neq}m'}d^{\dagger}_{m\up}d^{\dagger}_{m\dn}d_{m'\dn}d_{m'\up}
\end{multline}
The first three terms correspond to electron density-density ($n$) interactions: interactions between electrons with opposite spins in the same orbital ($U$), electrons with opposite spins in different orbitals ($U'$) and parallel spins in different orbitals ($U'-J_{\mathrm{H}}$). The two latter terms describe inter-orbital Hund's coupling interactions with electron creation and annihilation operators ($d$). The Hamiltonian for spin-orbit coupling is expressed as follows:
\begin{equation}
H_{\mathrm{SOC}} = \zeta~\sum\limits_{i}\vec{l_i}{\cdot}\vec{s_i}
\end{equation}
where $\zeta$ is the atomic spin-orbit coupling and $\vec{l_i}$ and $\vec{s_i}$ are the orbital and spin angular momenta vectors for individual electrons, respectively. The single-ion anisotropy for trigonal distortion is described as follows:
\begin{equation}
H_{\mathrm{tri}} = \Delta~\sum\limits_{i}({l^z})^2
\end{equation}
where $l^z$ refers to the $z$ component of the orbital angular momentum in the trigonal coordinate system. $\Delta$ is the trigonal crystal field and $\Delta < 0$ refers to compression. The total Hamiltonian is the sum of the above:
\begin{equation}
\label{eq:hamiltonian}
H_{\mathrm{total}} = H_{\mathrm{C}}+H_{\mathrm{SOC}}+H_{\mathrm{tri}}
\end{equation}
which is diagonalised in order to obtain the energy levels and multi-electron wavefunctions of Ru$^{4+}$ ion. The relevant electronic parameters were optimised so that the calculated energy levels match the observed RIXS peak positions. The comparison of obtained parameters for In$_2$Ru$_2$O$_7$ and Y$_2$Ru$_2$O$_7$ is given in Supplementary Table S5~\cite{Hozoi2014}.

\subsection{Muon spin rotation}
The Muon Spin Rotation ($\mu$SR) experiment on polycystalline In$_2$Ru$_2$O$_7$ was performed in the zero field (ZF) setup at the CHRONUS beamline at ISIS neutron and muon source~\cite{Tomono2010}. Spin polarised positive muons ($\mu^{+}$) were implanted in powder of In$_2$Ru$_2$O$_7$ enclosed in a silver foil packet. The ZF-$\mu$SR data were collected on warming from $\SI{10}{\kelvin}$ to $\SI{300}{\kelvin}$ in $\SI{10}{\kelvin}$ steps with 20-50 million muon events per scan and then normalised and binned using in-house WiMDA software~\cite{Pratt2000}. For detail on analysis, refer to the Supplementary Information.

\subsection{Band structure calculations}
We carried out band structure calculations based on the local density approximation (LDA) using the relativistic linear muffin-tin orbital (LMTO) method as implemented in the PY LMTO computer code~\cite{Antonov2004}. Spin-orbit coupling was accounted for by solving the four-component Dirac equation inside an atomic sphere which allows to easily calculate $J$ resolved densities of states (DOS) where $J = l \pm 1/2$ is the total angular momentum.

\section{Acknowledgements}
We are grateful to G. Khaliullin, J. Chaloupka, U. Wedig, G. Jackeli, A. V. Boris, T. I. Larkin, K. S. Rabinovich, G. M. McNally, P. van Aken, H. Walker and Y. Wang for fruitful discussions. We thank E. Buchner, S. Prill-Diemer, V. Duppel, F. Falkenberg, K. Schunke, U. Engelhardt and C. Stefani for experimental support. We acknowledge the provision of beam time for the neutron diffraction experiment on HRPD (Proposal No. RB1720369) as well as the provision of access to Materials Characterisation Laboratory facilities to the Science and Technology Facilities Council (STFC). The ZF-$\mu$SR experiment was performed on CHRONUS at ISIS Neutron and Muon Source with the approval of RIKEN-RAL muon facilities (Proposal No. RB1970108). The RIXS experiments were carried out at the beamline P01 of PETRA III at DESY. XAFS experiments were performed at BL14B1 of SPring-8 with the approval of the Japan Synchrotron Radiation Research Institute (JASRI) (Proposal No. 2020A3657). This work was partly supported by the Alexander von Humboldt foundation.

\section{Author contributions}

T.Ta. conceived the research. A.K. and M.I. synthesised the single-crystal and powder samples. A.K. performed transport, magnetization and thermodynamic measurements and the analysis. A.Y. performed electronic band structure calculations and the analysis. J.N. performed single crystal X-ray diffraction experiments and crystal structure solution. S.B. and R.D. performed X-ray powder diffraction experiments and the analysis. A.K., T.Ta. and A.S.G. performed neutron powder diffraction experiments and the analysis. J.B., H.G., A.K., M.B. and B.K. performed the RIXS experiment and A.K. and M.B. performed the analysis. A.K., T.Ta., D.P.S. and I.W. performed the $\mu$SR experiment and A.K. performed the analysis. D.M., T.Ts. and K.I. performed the EXAFS experiment and the analysis. H.T. supervised the project. A.K. and T.Ta. wrote the manuscript with input from all co-authors.

\section{Competing interests}

The authors declare no competing interests.

\end{document}